# A Metasemantic-Metapragmatic Framework for Taxonomizing Multimodal Communicative Alignment

Eugene Yu Ji[*]

[*]*yuji1@uchicago.edu*
*Cognitive Science, The University of Chicago, Chicago, IL, 60637, USA*

**Abstract**

Drawing on contemporary pragmatist philosophy and linguistic theories on cognition, meaning, and communication, this paper presents a dynamic, metasemantic-metapragmatic taxonomy for grounding and conceptualizing human-like multimodal communicative alignment. The framework is rooted in contemporary developments of the three basic communicative capacities identified by American logician and pragmatist philosopher Charles Sanders Peirce: iconic (sensory and perceptual qualities), indexical (contextual and sociocultural associations), and rule-like (symbolic and intuitive reasoning). Expanding on these developments, I introduce the concept of indexical contextualization and propose the principle of "contextualization directionality" for characterizing the crucial metapragmatic capacity for maintaining, navigating, or transitioning between semantic and pragmatic modes of multimodal communication. I contend that current cognitive-social computational and engineering methodologies disproportionately emphasize the semantic/metasemantic domain, overlooking the pivotal role of metapragmatic indexicality in traversing the semantic-pragmatic spectrum of communication. The framework's broader implications for intentionality, identity, affect, and ethics in within-modal and cross-modal human-machine alignment are also discussed.

## 1. Introduction

Multimodality is a significant area of research in psychology, cognitive science, and philosophy, and has recently gained critical importance in the rapidly evolving field of generative artificial intelligence (AI). However, the challenge of systematically understanding and evaluating multimodality in human cognitive models and contemporary generative Large Multimodal Models (LMMs) remains a notoriously complex task. While Large Language Models (LLMs), as LMMs' most prominent single-modal counterpart, still grapple with basic evaluative issues such as functional capacity and interpretability (Mahowald et al., 2024), LMMs introduce exponentially larger parameter spaces and evaluative dimensions, the complexity of which arises from the need to account for not just linguistic or any single non-linguistic domain alone, but the intricate ways in which different perceptual, cognitive and communicative modalities, such as text, image, and sound, sophisticatedly interact, co-construct, and generate meaning.

These challenges underscore the necessity of new interpretive, evaluative, and alignment methodologies for generative LMMs and, more generally, for modeling and comprehending human-like multimodal communicative capacities overall. The need for novel multimodal frameworks for many purposes, particularly that of benchmarking and evaluating human-machine alignment and misalignment, is not simply a technical requirement but also a conceptual and theoretical one. On the other hand, the fact that multimodality itself has been a longstanding subject of inquiry especially in psychology, cognitive science, and philosophy, offers a rich intellectual and scientific tradition from which multimodal AI research can draw. Broadly, this paper positions that methodologies developed in these disciplines, particularly in relation to theories of perception, cognition, context, and meaning can be adapted to create more comprehensive interpretive and evaluative frameworks for modeling human-like multimodal communicative capacities. While the boundaries between artificial and human intelligence in single modalities have been extensively redrawn or redefined by the recent development in LLMs (Blank, 2023; Shiffrin & Mitchell, 2023; Binz & Schulz, 2023), the characterization of multimodal intelligence in the AI literature remains, arguably, predominantly influenced by engineering-centric or at times folk-style conceptualizations of text, voice, image, video, motion, among others. This underscores the particular need for a systematic conceptual and philosophical foundation relating human and artificial intelligence of multimodality and aligning the latter with human multimodal capacities themselves (Ji, 2024).

In line with several recent critiques, I consider that current methodologies and evaluative frameworks for human and machine language modeling tend to overemphasize semantic-like and content-based reasoning at large, whether deductive or inductive reasoning, such as domains like symbolic capacity, common-sense reasoning, or daily life physics, which I collectively refer to as the *rule-like* modality (Pylyshyn, 1986)[1]. At the same time, a primary objective of current communicative LMMs is to align linguistic and perceptual inputs and outputs, such as in text-image, text-video, and text-voice models, where the main methodology often involves streamlining the mapping between non-rule-like, sensory-perceptual domains and rule-like ones. Further, although the sociocultural implications and impacts of within-modal and cross-modal models are widely discussed and recognized as critical concerns by many cognitive scientists and AI pioneers and practitioners (Griffiths, 2015; Raul et al., 2024; Crawford, 2021), they are often deemed as practical and applied problems only and there remains a lack of a robust and systematic framework for conceptualizing these practical issues along with the computational and modeling basis of cognitive models or generative AI. This gap leaves the full scope of natural and artificial models' multimodal capacities insufficiently understood or explored.

A key theoretical contribution of this paper is to highlight the importance of *metapragmatic capacity* as the most crucial mechanism for categorically navigating the relationships between semantic and pragmatic domains in linguistic and multimodal communication. To this purpose, I first introduce the Peircean framework, grounded in the semiotic theories that are first conceptualized by the 19th-century philosopher and logician Charles Sanders Peirce (1831 – 1914), and been since further developed in the contemporary "trading zones" (Galison, 1997) across philosophy, cognitive science, and linguistics (Silverstein, 1976, 1993; Jaszczolt 2016; Nakassis, 2018). At its foundation, the Peircean framework offers a systematic, human-grounded taxonomy for understanding how communicative modalities—semantic, pragmatic, sociocultural, and perceptual—relate, interact, and transition. I contend that developing a dynamic, meta-level taxonomic framework based on the Peircean iconic, indexical, and rule-like notions could facilitate the development of more effective interpretive methodologies and evaluative benchmarks for aligning intra-modality and inter-modality for both human and AI communicative models.

---

[1] As similarly used in, for instance, Pylyshyn (1986) and Greenberg (2023), the term "rule-like" in this paper does not necessarily imply rule-based or logic-driven processes. In logic, the statement like "a bear is an animal" is a necessary, categorical statement, while "all bears are dangerous" would be a probabilistic and contingent statement. Nevertheless, in natural intelligence, both examples can be interpreted as rule-like associations, with the second example perceived through the maximal qualifier 'all.' " Peirce collectively referred to such linguistic competence as "symbolic," emphasizing how a sign appears "symbolic" to an interpreter (Peirce, 1992, 1998); yet such a use only partially overlaps with the most common usage of symbolic representation in contemporary analytic philosophy and cognitive science. To avoid potential confusion, the term "rule-like" is used in this paper instead.

The remainder of the paper is organized as follows: Section Two introduces the basis of a meta-level, Peircean multimodal taxonomy. Section Three elaborates the importance of indexicality within this framework by developing the concept of indexical contextualization and the principle of "contextualization directionality." Section Four proposes a metasemantic- metapragmatic approach to semantic and pragmatic alignment, and examines its implications for within-modal and cross-modal alignment questions in intentionality, identity, affect, and ethics. The final section concludes the paper.

## 2. A Meta-level Multimodal Taxonomy

In recent scientific and AI literature, tensions between knowledge, communication, intramodal and intermodal capacities, and sociocultural sensitivity and impact have been widely acknowledged and debated (Raul et al., 2024). However, this tension has profound theoretical and historical roots in two classical dualisms that are shared across AI, philosophy, and cognitive science. The first contrasts semantic, structural, and symbolic knowledge with situational, contextual, and pragmatic knowledge (Putnam, 1975; Goldsmith & Laks, 2019), while the second contrasts the former types of knowledge with embodied, iconic, and perceptual ones (Barsalou, 1999, 2008; Lupyan & Winter, 2018; Greenberg, 2023). In the realm of AI, the challenge of incorporating situational, contextual, and pragmatic capacities, on the one hand, is widely discussed especially for intramodal text-based LLMs (Mahowald et al., 2024). On the other hand, embodied, iconic, and perceptual capacities are deemed essential for intermodal modeling, such as text-image and text-voice models, robotics, among others (Driess et al., 2023; Chia et al., 2024; Tang et al., 2024; Chen et al., 2024). These two conceptions give rise to distinct empirical methodologies for linking perception-related and cognition-like capacities with the broader world, and contextual and pragmatic aspects are rarely explored along with embodied and perceptual capacities in theoretical and empirical research — both approaches, however, grapple with similar challenges of bridging these capacities with semantic and rule-like modes of meaning.

In this section, I first propose a tripartite basis comprising iconic, indexical, and rule-like capacities. Building on this foundation, I develop a meta-level multimodal taxonomy aimed at bridging the above two seemingly incongruous dualisms in understanding and modeling meaning and communication.

**2.1 Iconic, Indexical, and Rule-like Capacities for Meaning and Communication**

I begin with the classical trichotomy of signs, icon, index, and symbol initially developed by Peirce (1992, 1998). Peirce's framework offers a systematic way to categorize the different ways in which meaning is constructed through domains from sensory-perceptual experiences to situational and sociocultural associations and logical or rule-like relationships. In this section, I break down how the Peircean categories of iconic, indexical, and rule-like can be adapted for understanding and modeling human-like multimodal capacities.

*- Iconic Capacities: Sensory and Perceptual Modes in Communication*

In the Peircean terminology, icons are signs that bear a direct resemblance or likeness to the objects they represent. Within the context of multimodal models, iconicity captures the sensory and perceptual properties that are central to how these models process inputs and generate outputs that are directly related to sensation and perception (vision, hearing, touching, tasting, smelling, sensorimotor, etc.). Specific iconic dimensions engage a cognitive system or a model's capacity to recognize or generate features that resemble or imitate the sensory-physical world, including images, sounds, textures, motion, among other sensory-perceptual phenomena.

From the Peircean perspective, what makes iconicity particularly intriguing is its relationship to specific and concrete *sensation and perceptually grounded* modes in communication. Intramodal and intermodal iconic elements in communication can invoke feelings or subjective experiences that are not strictly logical or factual, but are grounded in the sensory or environmental quality of human sensory and perceptual experiences. For example, a LMM that generates an image or an auditory output in response to a text prompt is leveraging iconicity to evoke specific sensory experiences (or even aesthetic or emotive responses, which will be further discussed in Section 4.3). In essence, the iconic mode of communication taps into a cognitive system or a model's capacity to process and learn inputs and generate outputs that are perceptually specific and evocatively resonant if necessary.

*- Indexical Capacities: Situational and Sociocultural Sensitivity*

The indexical category in Peirce's framework refers to signs that have a contextually associative connection to their referents. Particularly, indexical signs strongly relate to significant or specific socioculturally habituated associations that a human or model can recognize, replicate, or, in some cases, negotiate, manipulate, or deny. These associations are not based on perceptual or purely logical relationships per se, but are deeply embedded in situation and context, often predominantly associated with social practices and cultural meanings and dynamics.

Indexicality highlights a cognitive system or a model's ability to relate to *situationally and socioculturally contiguous connotations* and to recognize the significance of such connotations (Silverstein, 1993), such as when a human or model is tasked with generating language or images appropriate for a specific sociocultural setting or particular group of human or agentive population. This extends to recognizing specific idioms, visual aesthetics linked to certain social groups, or stylistic preferences that signal identity, status, or affiliation, among others. Empirically, evaluating a cognitive system or a communicative model on its indexical capacity thus involves assessing its performance in recognizing the significance of specific sociocultural markers and adapting to sociocultural nuances and conventions. On the other hand, indexicality in this paper is conceived as much more than just a domain that imposes sociocultural connotations on iconic or rule-like processes (See Section 3).

- *Rule-like Capacities: Symbolic and Intuitive Reasoning at Large*

The rule-like mode, which can be approximately mapped onto the Peircean notion of symbolic representation, encompasses signs interpreted as factual, truthful, rule-like, following any types of general principles or generalizing reasoning. I characterize rule-like capacities as involving assessing how well a cognitive or computational system adheres to factual accuracy, processes complex linguistic structures or mathematical relationships, applies classical induction and deduction, and navigates scenarios requiring an understanding of law-like regularities in the everyday world. In most contemporary work in analytic philosophy, cognitive science, and computer science, these domains are typically considered in a cross-domain manner. Recent developments have raised questions about, for example, why large generative models seem to handle semantic summarization more readily than arithmetic reasoning (Xu et al., 2024), or whether AI should grasp intuitive physics rather than merely extracting meaning from statistical co-occurrence patterns in text or pixels (LeCun, 2022). From the Peircean perspective, however, an important common characteristic shared across these cases is certain perceived or perceivable intrinsic regularity or internal coherence, whether within a sentence, an equation, a logical sequence, or an image. Regardless of whether these patterns emerge from logical procedures, symbolic representations, or statistical regularities, they exhibit a cognitive quality that is neither purely sensory-perceptual nor contextually associated, but *appears in communication* t*o be inherently rule-like*, characterized by a certain level of within-domain or cross-domain consistency and coherence.

In the context of aligning human or human-like multimodal modeling, I use rule-like capacities to refer to the capacity to engage in *symbolic and intuitive reasoning at large* and in a broad sense. This involves a human or model's capacity to apply rule-like (but *not* necessarily rule-based) processes of meaning-making, driven by internal regularities, such as making predictions based on laws of physics or general rules, commonly accepted facts, or truths and intuitions about the world. For example, a LMM that can infer generic causal or correlative relationships, reason about physical interactions such as objects falling due to the gravitational law, track the truth conditions of a sentence, or evaluate the coherence of a narrative or an image is drawing upon its rule-like capacity.²

## 2.2. Meta-level Communicative Taxonomy Generated from the Three Basic Capacities

While the three basic capacities of communication are fundamental, they are not ontologically static, but epistemologically and communicatively generative across the entire spectrum of human communication. In realistic communication, humans often reflectively interpret images as carrying specific cultural implications, treat specific social conventions as generalizable or rule-like, or imbue objective facts with communal or personal connotations. Much existing work in AI tends to predominantly treat many of these domains as origins of misalignment, biases, or issues to be mitigated, but they reflect essential aspects of human behavioral and communicative spectrum. The rich *meta-level* interpretative capacities *superimposing* the three basic communicative modes must be systematically incorporated in the context of aligning generative multimodal models for both human and machine.

Specifically, there are three possible *downshifting* meta-level superimposing processes, where interpretive transitions occur from rule-like regularities to socioculturally conventional or perceptual regularities: *from rule-like to iconic*, *from rule-like to indexical*, and *from indexical to iconic*. Correspondingly, there are three *upshifting* meta-level superimposing processes, which move in the opposite direction: *from iconic to indexical*, *from iconic to rule-like*, and *from indexical to rule-like*. While Peirce offers a partial theoretical account of these meta-level, second-order multimodal relationships (Peirce, 1992, 1998), and contemporary sociolinguistics and linguistic anthropology further develop numerous theoretical and empirical studies on the interplay among iconic, indexical, and rule-like communicative modes (Silverstein, 2003; Ball, 2014; Gal & Irvine, 2019; Nakassis, 2023), a comprehensive meta-level taxonomic framework tailored specifically for the multimodality questions in human-human and human-machine communication remains largely undeveloped. In Table 1, I demonstrate how a comprehensive Peircean taxonomic framework accommodates both basic (1st order, non-meta-level) and superimposing (2nd order, meta-level) modes for multimodal communication.

---

² Work in the history of science and philosophy reveals that terms like "laws" and "rules" have often been used interchangeably across various traditions and cultures (Daston, 2022). The distinction between law-like, deterministic, and causal cognition on one hand, and regularity-driven, probabilistic, and statistical cognition on the other, only gained prominence in modern Western philosophy during the post-Cartesian era, and cognitive science and AI inherited this binary framework at their inception (Goldsmith & Laks, 2019). Rather than adhering to this particular dualistic approach, this paper treats both together under the umbrella term "rule-like" and adopts an alternative, tripartite framework, contrasting or relating the rule-like mode of communication with iconic and indexical ones instead, aiming to shed new light on these longstanding debates.

|  | **Taxonomy of Multimodal Communicative Capacities** |
|---|---|
| Non-meta level (First-order Bases) | 1. Iconic:<br>    *e.g., The color is bloody.*<br>2. Indexical:<br>    *e.g., The city hall truly represents this city.*<br>3. Rule-like:<br>    *e.g., The figure looks upside down.* |
| Meta-level (Second-order Superimposition) | 4. Rule-like → Indexical (metapragmatic; downshifting):<br>    *e.g., The upside-down figure looks sci-fi.*<br>5. Rule-like → Iconic (metasemantic; downshifting):<br>    *e.g., The unrecognizable handwritten scripts feel grotesque.*<br>6. Indexical → Iconic (metapragmatic; downshifting):<br>    *e.g., The wedding cake is so sweet and delicious.*<br>7. Iconic → Indexical (metapragmatic; upshifting):<br>    *e.g., This colorful attire is as colorful as Thai culture.*<br>8. Iconic → Rule-like (metasemantic; upshifting):<br>    *e.g., Winter in Chicago is always white and windy.*<br>9. Indexical → Rule-like (metapragmatic; upshifting):<br>    *e.g., Everyone wears formally in a wedding.* |

**Table 1.** Taxonomy of multimodal communicative capacities divided into a tripartite basis and six meta-level, second-order possibilities superimposed upon the basis.

At the non-meta, first-order level, the iconic mode refers to perceptual and sensory properties or regularities, as demonstrated in the example "The color is bloody," which captures the immediate, perceptual resemblance inherent in iconic signs. The indexical mode involves socio-cultural or contextual associations, as in the example "The city hall truly represents this city," where a landmark site indexes a specific geographic and cultural context. The rule-like mode, on the other hand, encompasses rule-like regularities, such as "The figure looks upside down," where an observed visual feature is interpreted *as if* it is logically or unconditionally true or false.

The six meta-level, second-order processes account for the dynamic superimpositions across the three basic, first-order modes of communication. Downshifting transitions occur when meaning moves from more rule-like regularities toward indexical or iconic interpretations. For instance, in the transition from rule-like to indexical, as in "The upside-down figure looks sci-fi," a seemingly unconditional feature ("upside-down") is imbued with a genre-based, sociocultural association ("sci-fi"). Similarly, in the transition from rule-like to iconic, as in "The unrecognizable handwritten scripts feel grotesque," rule-like irregularity ("unrecognizable script") generates a perceptual and affective response ("grotesque"). Another downshifting example is the indexical to iconic transition, where socioculturally contextual associations produce perceptual interpretations, as in "The wedding cake is so sweet and delicious," linking the sociocultural context ("wedding") to strengthen a sensory experience ("taste").

Conversely, meta-level upshifting transitions move in the opposite direction, from more perceptual or contextual interpretations toward rule-like regularities. In the shift from iconic to indexical, as in "This colorful attire is as colorful as Thai culture" a perceptual attribute (colorfulness) is interpreted within a broader cultural context. The transition from iconic to rule-like, exemplified by "Winter in Chicago is always white and windy," moves from sensory perception ("white and windy") to a generalized, seemingly truth condition-like understanding of a seasonal pattern in a geographical location. Finally, the shift from indexical to rule-like is illustrated by "Everyone wears formal attire at a wedding," where a sociocultural norm ("formal attire") becomes codified into a rule-like regularity.

Thus, tripartite multimodal bases and six meta-level superimposed multimodal modes of communication are identified. In the next section, I propose and elaborate on the argument that indexical contextualization serves as the central channeling and mediating meta-level mechanism across the tripartite multimodal bases.

# 3. Indexical Contextualization as the Central Metapragmatic Capacity to Navigate across Communicative Modes

How does human cognition effectively navigate the vast array of interpretive possibilities across the diverse non-meta and meta-level multimodal modes in communication? In this section, I propose that *indexical contextualization* (Silverstein, 1976, 1993; Gumperz, 1982) can serve as a key *metapragmatic* mechanism for reflectively navigating and managing the shifts between indexical and non-indexical (rule-like and iconic) communicative modes (components 4, 6, 7, 9 in Table 1). This idea draws from both contemporary pragmatics theories and philosophy of language, where indexical contextualization is central to the theorization of metapragmatics (Silverstein, 1993; Urban, 2006). A central theoretical proposal in this work is that metapragmatic awareness, facilitated by indexical contextualization, is crucial for enabling cognition to maintain or adapt communicative forms and norms across varying situational and sociocultural contexts, allowing for a dynamic, fluid, and empirically grounded interpretation of meaning in context.

While current cognitive and LMM modeling does not entirely neglect switching between communicative modes, it disproportionately focuses on rule-like relations and rule-like *metasemantic* superimpositions (*e.g.*, components 5 and 8 in Table 1). Although contextual and sociocultural connotations and implications are widely acknowledged as essential in contemporary approaches to modeling pragmatic reasoning, they are typically addressed either through limited inputs from pragmatic theories, most commonly the Rational Speech Act (RSA) framework (Grice, 1975; Degen, 2023; for computational models based on RSA, see Goodman & Frank, 2016; Frank, 2016; Erk, 2022), or in an ad-hoc or post-hoc manner that often ineffectively approximates metapragmatic awareness of rule-like expressions. Similarly, sensory and perceptual dimensions are also often modeled according to rule-like criteria, most typically as representational capacities in sensation and perception (Williams & Colling,

2018; Springle, 2019), rather than more broadly, dynamically defined iconic capacities[3]. On the other hand, in contemporary cognitive science and philosophy of mind, both rule-like and iconic capacities have been intensively theorized and investigated, and there has been extensive work arguably addressing superimposing certain iconic relations as rule-like ones, such as under the framework of embodied or grounded cognition (Varela et al., 2016; Barsalou, 1999, 2008), or work developing the reverse process laminating rule-like relations with specific iconic ones, such as metaphorical reasoning and cross-domain mapping (Lakoff, 1993). However, in these works, cognitive and communicative processes that should be recognized as indexical and metapragmatic are often improperly simplified or approximated as semantic or metasemantic, limited within rule-like and iconic dimensions only.

**3.1 Three Types of Indexical Contextualization**

Extended from the foundational work of Silverstein (1976, 1993), I identify three types of indexical contextualization—*encontextualization*, *decontextualization*, and *recontextualization*—which together form a cohesive framework for modeling how meaning is metapragmatically mediated across basic and meta-level communicative modes:

1. Encontextualization refers to the process by which iconic or rule-like meanings are saliently embedded within specific indexical contexts. This process involves registering and aligning abstract or perceptual meanings with socially or culturally situated interpretations. For instance, a visual representation (iconic) might be encontextualized as an index of political resistance when situated within a particular social or cultural context, thereby aligning appropriated perceptual or sensory properties with specific sociocultural connotations.

---

[3] Research on dynamic rather than representational, often visually dominated, iconicity is evident in contemporary studies within the philosophy of mind (Dreyfus, 2007), philosophy of perception (Barwich et al., 2024), and perceptual neuroscience (Freeman, 2008; Kay, 2003). While extensions of iconicity in this vein are beyond the central scope of this paper, they will be further incorporated and developed in future work.

2. Decontextualization involves the selective softening, inhibition, or removal of an encontextualized socio-cultural connotation. This can occur when a specific iconic or rule-like meaning needs to be less bound by its original context to be made interpretable in alternative ways. For example, a culturally loaded image might be decontextualized to be interpreted as an indexically neutral or ambivalent sign, distanced from the original specific indexical connotation.
3. Recontextualization is the process of realigning the expression to fit an alternative indexical possibility. This often occurs when a particular sign is successfully re-adapted or reframed to suit a new indexical context. For instance, a gesture that carries a specific indexical meaning in one context might be recontextualized to fit a different communicative scenario, thereby altering its indexical associations.

In communication, encontextualization channels what is contextually and indexically salient. Decontextualization is the opposite process of encontextualization, channeling communication to be open and flexible to alternative contextual and associative possibilities. Recontextualization can be understood as the attempt to pursue and secure at least one alternative way of encontextualization following from the prior process of decontextualization, allowing communicators to indexically adjust to align with uncertain, shifting, or new understanding of situations and contexts. The three contextualization capacities for managing contextual maintenance or transitions at the metapragmatic level are especially relevant in comprehending and modeling human and human-machine multimodal interactions, where multimodal generative and interactional prompts often require both context-sensitive precision and context-malleable adaptability across linguistic, sociocultural, and sensory-perceptual modalities.

**3.2 Principle of Contextualization Directionality in Socio-ecological Communication**

The three types of contextualization vary in their prevalence across communicative contexts. Particularly, for communication that becomes *socio-ecological*, *i.e.*, a communicative scenario involves more than one interactional pair and becomes chain-like or network-like in the socio-ecological environment[4], I propose that indexical contextualization should adhere to *The Principle of Contextualization Directionality*:

- Principle of Contextualization Directionality: Within a socio-ecologically relatable communicative scenario, once a communicative element has been indexically *en*contextualized, aligning it with a specific contextual milieu, its immediate *de*contextualization becomes increasingly challenging: any attempt of decontextualization, *i.e.*, to alter or remove the previous encontextualization, would motivate socio-ecologically proper, subsequent *re*contextualization to ensure the metapragmatic coherence and integrity of the entire communicative scenario[5].

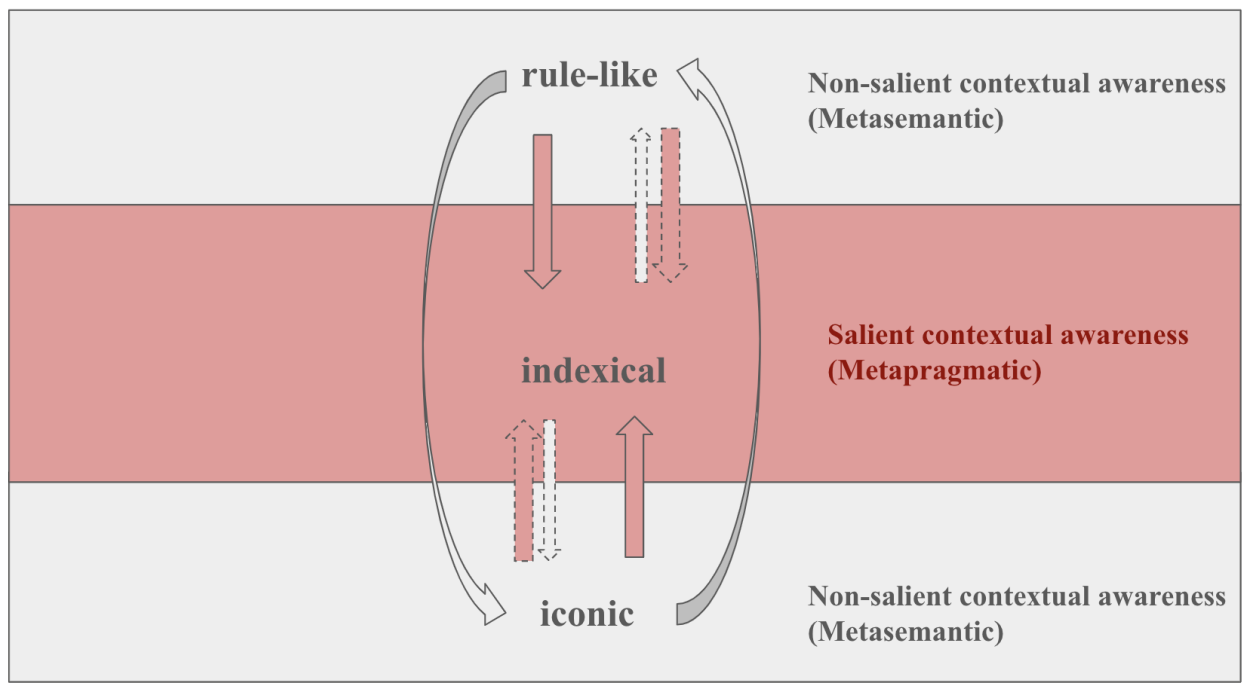

**Figure 1**. Principle of Contextualization Directionality: In the same or across socio-ecologically relatable communicative scenes, once indexical encontextualization (light red arrows) is performed, decontextualization (dotted gray arrows) becomes increasingly difficult unless a proper recontextualization (dotted light red arrows) is expected to be plausible. In contrast, transitions between rule-like and iconic modes (curved gray arrows) do not exhibit asymmetrical directionality, as both modes indicate similarly non-salient contextual awareness.

---

[4] As discussed, for instance, in Goffman's "interaction order" (1983), Gumperz's "interactional sociolinguistics" (1982), and Collins's "interaction ritual chain" (2004).

[5] Put in a broad context of pragmatic theory and modeling, I posit that The Principle of Contextualization Directionality represents a more fundamental maxim than cooperative principles such as Grice's Cooperative Principle (1975) and usual coordination-based approaches, as contextualization directionality provides the essential foundation for navigating and transitioning between semantic and pragmatic modes in communication as a prerequisite for cooperation or coordination to emerge within or across either communicative mode.

Figure 1 illustrates the "communicative flows" dictated by contextualization directionality across rule-like, indexical, and iconic modes of communication, as performed by the metapragmatic processes of indexical contextualization. As shown in the figure, indexicality occupies a central meta-level mediating role, with asymmetry between the transition from rule-like or iconic modes to the indexical mode and the transition from the indexical mode back to rule-like or iconic modes. In other words, once a communicative element undergoes indexical encontextualization (light red arrows), where it is aligned with specific contextually sensitive setting or associated with salient contextual awareness at the meta level, it becomes significantly more challenging, within the same or socio-ecologically related communicative scenario, to reverse this process or inhibit the contextual awareness to move back to a first-order rule-like or iconic mode with non-salient contextual awareness (dotted gray arrows). On the other hand, decontextualization is more likely to occur if a proper recontextualization is performed or anticipated to be performed in communication (dotted light red arrows). For example, in a conversational scene where an iconic image is encontextualized with specific political connotations, it would be difficult to strip those connotations away. *i.e.*, to "de-politicize" them and present or narrate the image as predominantly "perceptual" or "semantic." Such a shift would require proper recontextualization, where the updated context sufficiently justifies the removal or alteration of the original indexical framing, or a return to it. Without this, a decontextualization process is unlikely or would become ineffective, leaving the original encontextualized connotations to continue to influence interpretation.

Meta-level transitions between rule-like and iconic modalities without explicit indexical mediation, on the other hand, allow for pragmatically more bi-directional or even symmetrical shifts (curved gray arrows in Figure 1). This does *not* imply that rule-like and iconic modes are devoid of indexicalizable socio-cultural connotations, but rather that, in a communicative event, such connotations need to be explicitly evoked at the metapragmatic level by indexical superimposition (*i.e.*, through encontextualization). In many cases, however, specific possibilities for encontextualization either fail to become salient or remain sufficiently stable and consistent that they go unnoticed by the communicators. For example, when two mathematician friends seamlessly collaborate on a math problem, or when an art student sketches a natural scene willingly following a simple pedagogical prompt like "drawing a bird on a beautiful tree" from an art teacher, the immediate socio-cultural connotations often remain implicit and unexamined.[6]

The next section will further propose a metasemantic-metapragmatic framework of dynamic, multimodal communicative alignment and examine the broader theoretical and empirical implications of The Principle of Contextualization Directionality embedded in this framework.

## 4. Metapragmatic-centric and Metasemantic-centric Alignments
### 4.1 Alignment, Misalignment, Realignment via Contextualization

Following the formulations on directionality in contextualization in the previous section, I identify two main types of alignment problems. The first type involves salient contextualization, focusing on communicative dynamics enabled by metapragmatic awareness that entails transitions to or from indexicalized communicative elements, and alignment issues of this type require careful attention to the directionality of contextualization. The second type does not explicitly involve salient contextualization and contextualization directionality, primarily dealing instead with not saliently indexicalized or contextual, metasemantic relations within or between iconic and rule-like elements.

For the metapragmatic alignment problem, consider the following example: Alice (a human or communicative machine) says to Bob (another human or communicative machine), "The weather is nice today," the encontextualization of which forms a typical way to initiate a conversation in many culturally salient contexts, using weather as a socially acceptable indexical cue to break the ice. If Bob then asks "Why are you talking about the weather?", this would form a step of decontextualization, where the initial encontextualized meaning is questioned or challenged, distancing it from Alice's initially assumed or intended context. Alice could respond, "Oh, I thought it was a nice icebreaker at a banquet," thus providing a recontextualization that explicitly explains the original indexical meaning and re-establishes the contextual significance of the weather comment. Bob then could acknowledge with, "Oh, yep! Nice weather and nice banquet!" completing the recontextualization through eventually aligning with Alice's original encontextualized context, affirming the shared understanding of the communicative context indexicalized by weather.

---

[6] In instances where an iconic or rule-like element does not evoke salient contextual awareness, these elements can be positioned along what Greenberg (2023) terms the "iconic-symbolic" spectrum. On the other hand, it should be noticed that non-salient contextual awareness does not imply that an iconic or rule-like element is devoid of context or any possibility of contextualization. It simply means that it is not saliently encontextualized in a particular moment within a communicative scene. For humans, the cognitive repertoire for encontextualization and recontextualization is rich and diverse, but whether they are evoked in communication in one way or another depends on the context, identities, and agency involved in a specific interactional scenario.

However, consider an alternative scenario: Alice says, "The weather is nice today," Bob asks, "Why are you talking about the weather?", and Alice similarly replies, "Oh, I thought it was a nice icebreaker." But if Bob responds with, "Yes, there are no clouds out there," this response fails to successfully recontextualize the exchange, as Bob misses the opportunity to address the intended alignment and instead focuses on the literal (iconic) but not contextual (indexical) comments on weather conditions. In human conversations, such failures of recontextualization, if they occur, are often perceived as socially misaligned and problematic. This is because successful decontextualization typically requires an effective recontextualization (either by realigning with the existing context or establishing a new context) to maintain pragmatic coherence in communication. Once a communicative element is encontextualized in a particular way, it becomes increasingly challenging or effortful to decontextualize or shift its meaning into a different communicative mode (*e.g.*, from socio-cultural and indexical to seemingly purely rule-like or iconic) without appropriate recontextualization. Typically, when an attempt of encontextualization is understood by the other party, decontextualizing words or questions, like Bob's "Why are you talking about the weather?" would not be raised, if both sides anticipate and are willing to coordinately avoid potential misalignments or communicative breakdowns that might follow potential decontextualization.

|  | Examples of Alignment | Examples of Misalignment |
|---|---|---|
| **Metapragmatic (Mis-)Alignment under salient contextual awareness** | A: "The weather is nice." (encontextualization)<br><br>B: "Why are you talking about the weather?" (decontextualization)<br><br>A: "Oh, I thought it was a nice icebreaker at a banquet." (recontextualization)<br><br>B: "Oh, yep! Nice weather and nice banquet!" (alignment through recontextualization) | A: "The weather is nice." (encontextualization)<br><br>B: "Why are you talking about the weather?" (decontextualization)<br><br>A: "Oh, I thought it was a nice icebreaker." (recontextualization)<br><br>B: "Yes, there are no clouds out there." * (misalignment and failure in recontextualization) |
| **Metasemantic (Mis-)Alignment under non-salient contextual awareness** | A: "The weather is nice. There are no clouds today." (iconic/descriptive)<br><br>B: "Yes, the summer on the beach is always like that." (rule-like/factual alignment)<br><br>- *Alternatively:*<br>A: "The summer on the beach is always like that." (rule-like/factual)<br><br>B: "Yes, the weather is nice. There are no clouds today." (iconic/descriptive alignment) | A: "The weather is nice." (iconic/descriptive)<br><br>B: "Yes, it's raining." * (iconic/descriptive misalignment)<br><br> - *Alternatively:*<br> "No, the weather is not nice." * (rule-like/factual misalignment) |

**Table 2.** Examples of metapragmatic and metasemantic alignment and misalignment, applicable to both intramodal and intermodal scenarios. Interlocutors ("A" = Alice and "B" = Bob) can be either humans or communicative agents.

The metasemantic alignment problems, on the other hand, are not particularly sensitive to directionality in transitioning between communicative modes. Consider that if Alice says to Bob, "The weather is nice. There are no clouds today." (intended by Alice as a descriptive, iconic expression). Bob responds, "Yes, the summer on the beach is always like that" (a transition from iconic to rule-like). In this case, the communication is successful as both parties maintain metasemantic coherence between iconic and rule-like expressions. Different from the previous example where contextual awareness is salient and metapragmatic directionality matters, one can easily imagine this conversation proceeding in the reverse direction, from rule-like to iconic: for example, Bob says, "The summer on the beach is always like that" (rule-like), and Alice responds, "Yes, the weather is nice, there are no clouds today" (a descriptive/iconic response).

In metasemantic-driven exchanges, misalignment can occur only if factual mistakes, deductive or inductive contradictions, or violations of daily physics common sense arise. For example, Alice says, "The weather is nice," and Bob responds, "Yes, it's raining." This could indicate either a factual or knowledge error, or a shift in the conversation to an encontextualized mode (such as sarcasm or humor). Alternatively, Alice could say, "The weather is nice," and Bob might reply, "No, it's not." In such cases, Alice and Bob may engage in a debate over the factual or subjective nature of the statement. However, unless the conversation shifts into the alternative metapragmatic framing through recontextualization (such as about, again, sarcasm or humor), Any misalignment should ideally be realigned through consistent factual or descriptive expressions, reversible across non-saliently contextual directions. Alignments and misalignments of the metasemantic type, therefore, should be viewed differently from those of the metapragmatic one. Hence, the most immediate cognitive and computational strategies to resolve possible misalignments of these two types are also different: resolving contextualized and metapragmatic misalignments requires directionally oriented recontextualization processes, such as reaffirming the original context or effectively transitioning to a new context, rather than immediately decontextualizing the misalignment and reducing it to the iconic or rule-like realm. In contrary, non-contextualized misalignments, such as those involving descriptive, factual, or logical discrepancies, can be resolved through verifiable information, shared understanding, logical clarification, etc., or iterative exchanges among any of them, unless a possibility of encontextualization, intended or unintended, emerges during the communication.

Although the examples above specifically demonstrate intramodal scenarios of verbal communication, the same framework is readily applicable to intermodal and multimodal cases, the research direction of which is also explored in recent theoretical and empirical work within linguistic anthropology and sociolinguistics (Nakassis, 2023). Consider, for instance, a painting depicting foggy weather. A viewer or agent might descriptively assess its iconic elements, judging whether it accurately represents how foggy weather typically appears in the real world. However, once the painting is encontextualized as belonging to, for instance, the "impressionist" style, pursuing the same iconic or rule-like judgments only may lose relevance unless they adhere to that overall context (*e.g.*, interpreting hazy, blurry brushwork as a characteristic index of impressionism) or aim for de/recontextualization (*e.g.*, suggesting that the blurry style may resemble a myopic view as if seen without proper glasses).

Just as verbal or textual elements can be indexically contextualized, so too can images, pictures, sounds, among many other domains undergo similar processes under the same maxim of contextualization: The Principle of Contextual Directionality highlights the meta-level asymmetrical tendencies in navigating transitions between saliently indexical and non-saliently indexical communicative modes in both within-modal and multimodal interactions.

**4.2 Dynamic Alternations Between Metasemantic-Centric and Metapragmatic-Centric Alignment**

Thus, I extend the classical duality of metasemantics and metapragmatics in theoretical linguistics and the philosophy of language (Silverstein, 1993; Urban, 2006; Bublitz & Hubler, 2007; Jaszczolt, 2016, 2021) to a multimodal duality. Iconicity and rule-like capacities, by default (*i.e.*, in the absence of indexical superimposition), are characterized as non-indexical, semantic or metasemantic, which lack salient contextual awareness and rely primarily on perceptual, conceptual, or structural regularities. In contrast, indexicality is inherently salient at the metapragmatic level, capable of spotlighting any communicative scenario contextually aware through indexical superimposition. The Principle of Contextualization Directionality guides and shapes the meta-level possibilities of alternating between semantic and pragmatic modes as two interpretive orientations in communication.

This extended multimodal duality allows us to define a dynamic taxonomy with two alternative meta-level interpretive centers and hierarchies for multimodal alignment, in which indexicality can be defined as the alternative metapragmatic-driven interpretive center, while iconicity and rule-like regularities are defined as the metasemantic-driven interpretive center[7]:

- *Metapragmatic-Centric Alignment*: Indexicality has high salience due to its high contextual embeddedness, while iconic and rule-based elements exhibit low salience.

    Metapragmatic-Centric Alignment Hierarchy: iconic (low salience), indexical (high salience), rule-based (low salience)

- *Metasemantic-Centric Alignment*: Iconic or rule-based elements are seen as high in salience, while indexical elements hold a lower salience.

    Metasemantic-Centric Alignment Hierarchy: iconic (high salience), indexical (low salience), rule-based (high salience)

These two meta-level interpretive hierarchies operate in a discrete, categorical manner, alternating between one another in communication, the process of which follows the requirement of contextualization directionality. On the other hand, either of them can dynamically navigate communication across the entire semantic-pragmatic spectrum, allowing rich and highly flexible transitions between diverse within-modal and cross-modal communicative repertoires.

---

[7] Here, I consider the shared metasemantic nature of iconicity and rule-like regularities as approximately equivalent to metasemantic awareness of the iconic-symbolic spectrum in Greenberg (2023). Empirically, such a treatment also finds remarkable parallels in contemporary work across computational linguistics and computer vision, such as language-driven recurrent neural networks (RNNs) processing images (Gregor et al., 2015) and vision-driven convolutional neural networks CNNs generating syntactic and semantic outputs (Kim, 2014; Yin et al., 2017). A key objective of this paper is to treat the iconic-symbolic spectrum, as Greenberg (2023) puts it, as a unified *metasemantic* unit, in contrast to indexical contextualization as the *metapragmatic* focal point. Recent theoretical advancements in sociolinguistics and linguistic anthropology, nonetheless, begin to unveil the nuances within the metapragmatic relationship between indexicality and iconicity (Gal, 2013; Ball, 2014; Nakassis, 2023), and future work is required to expand in this direction.

The dual interpretive focuses highlight a dialectical relationship between metasemantics and metapragmatics (Silverstein, 1993) in multimodal scenarios. While semantic expressions (both rule-like and iconic ones) *may appear* non-pragmatic and not saliently contextual, the pragmatic connotations that a semantic-oriented expression or presentation carry can in principle *always* be rendered metapragmatic and saliently contextualized. On the other hand, all pragmatic expressions and metapragmatic awareness must ultimately be metasemantically and epistemologically *grounded* in knowledge, well-formedness, or reality, *i.e.*, attain foundationalist (McGrew, 2003; van Cleve, 2005), coherentist (Hage, 2013), or realist significance (Dreyfus & Taylor, 2015). Our framework delineates how multimodal communication can be oriented either metasemantically or metapragmatically, while preserving the diverse, fluid possibilities of interpreting icons, indexes, and rule-based elements over the entire semantic-pragmatic spectrum. A particular emphasis is paid, nevertheless, on the indexicality-driven, metapragmatic-centric alignment, as compared to the metasemantic-centric alignment, the former remain much more underdeveloped in the existing literature in philosophy, cognitive science, and computer science about communication.

## 4.3 Human-Machine Alignment of Intentionality, Identity, Affect, and Ethics via the Metasemantic-Metapragmatic Taxonomy

In this section, I discuss four empirical domains where I deem that indexicality and metapragmatics should be integrated: intentionality, identity, affect, and ethics, which are deemed crucial in both classical and contemporary literature related to human-machine multimodal alignment questions. Classical discussions on AI in philosophical and cognitive science usually focus on whether intentionality could be simulated or approximated in artificial systems (Dennett, 1987; Searle, 1980; Dreyfus, 2007). Related research specifically focused on what is often termed alignment problems today highlights the need to understand not only the explicit content of communication but also the underlying intentions that shape communication and interpretation (Stalnaker, 1978; Grice, 1989). This is particularly relevant in both single-modal and multimodal alignment where explicit or implicit, perceived or conceived intent can shift meaning significantly – according to our framework, such intent shift can occur between the metasemantic-centric and metapragmatic-centric modes, which is necessarily mediated via indexicality. Similarly, agentive and sociocultural identity is also widely discussed in generative large-model alignment (Lu et al., 2022; Tao et al., 2022; Tadimalla & Maher, 2024) As individuals necessarily bring contextually and usually socioculturally specific identities into communicative acts. According to our proposed framework, cognitive or AI models must be capable of identifying and adapting to both semantic and pragmatic identity markers, particularly differentiating them via their indexical values at the metapragmatic level (Silverstein, 1976; Eckert, 2000; Bucholtz & Hall, 2005) to effectively aligning socio-ecological connotations of multimodal outputs (Hovy & Spruit, 2016). Affect, or emotional alignment, represents another critical domain in multimodal alignment. Modeling work in both cognitive and computational sciences has long emphasized the importance of recognizing and generating outputs sensitive to emotional content, context, and affective cues (Picard, 1997; Calvo & D'Mello, 2010). Misaligned affective cues and outputs, whether within or across modalities, can result in significant communicative misalignment, and according to the metasemantic-metapragmatic framework, these issues cannot be fully resolved without incorporating indexical contextualization to account for how affect and emotion are conveyed in communication, alternative to or laminated with iconic or rule-like dimensions of affect such as the physiologically grounded basic and complex emotions (Ekman, 1992; Russell, 2003; Berrios, 2019). Last but certainly not least, ethical connotations, especially concerning justice, fairness, biases, stereotypes, among others, are increasingly foregrounded in research of cognitive modeling and AI alignment with respect to both single-modal and cross-modal scenarios, yet they are mostly conceptualized under either normative or descriptive ethics (Binns, 2018; Greene, Hoffmann, & Stark, 2019; Gebru et al., 2018; Danaher, 2020; Gabriel, 2022). The literature of sociolinguistics and linguistic anthropology has long considered ethical questions in society and culture as richly and sophisticatedly indexicalized (Silverstein, 2003; Agha, 2003; Eckert, 2008; Gal, 2013). This body of work can offer a critical empirical foundation for further theorizing how the metasemantic-metapragmatic duality mediates metaethical frameworks (Stevenson, 1944; Hare, 1982), bridging normative or descriptive ethics with pragmatic ethics (Dewey, 1982 [1920]; Bernstein, 1983; Margolis, 2007 [1986]; Liszka, 2021). Such an integration can provide an invaluable pathway for more effectively integrating ethical alignment into complex multimodal human-machine interactions (Gabriel, 2022; Kasirzadeh & Gabriel, 2023).

Together, intentionality, identity, affect, and ethics constitute a comprehensive challenge within the current empirical literature on multimodal human-related and human-machine communicative alignment. The challenge underscores the importance of a meta-level grounded framework that can navigate the complex content-context, semantic-pragmatic, regularity-contingency terrain across these four empirical domains in intramodal and intermodal settings, where metapragmatic- centric alignment necessarily forms a *dynamic* and *dialectic* counterpart to metasemantic-centric alignment, as demonstrated in Table 3:

|  | **Metasemantic-centric alignment**<br>- Iconically or rule-like oriented;<br>- Yet indexicable at the *metapragmatic* level | **Metapragmatic-centric alignment**<br>- Indexically oriented;<br>- Yet interpretable as iconic or rule-like at the *metasemantic* level |
|---|---|---|
| **Intentionality** | Content-oriented:<br>　*e.g.*, information, mental models, Theory of Mind. | Context-oriented:<br>　- Contextualization Directionality navigates how specific understandings of selves and/or others' intentions are evoked or suppressed in particular socio- ecologically sensitive contexts. |
| **Identity** | Demographic, socio-biological:<br>　*e.g.*, race, gender, age, name, income, cultural background, family ties, facial appearance, voice, etc. | Situationally, indexically evoked:<br>　- Contextualization Directionality navigates how specific sets of demographic and socio-biological dimensions of identity are evoked or suppressed in particular socio- ecologically sensitive contexts. |
| **Affect** | Physiologically grounded:<br>　*e.g.*, basic emotions & complex emotions | Situationally, indexically evoked:<br>　- Contextualization Directionality navigates how specific sets of physiologically grounded affective and emotional dimensions are evoked or suppressed in particular socio- ecologically sensitive contexts. |

| **Ethics** | Normative or descriptive: *e.g.*, moral principles, ethical standards, moral reasoning, ethics of public policies and rules, etc. | Pragmatist: <br> - Contextualization Directionality navigates how specific sets of moral principles, reasoning, or psychological responses are evoked or suppressed in particular socio-ecologically sensitive contexts. |
|---|---|---|

**Table 3.** Across the domains of intentionality, identity, affect, and ethics, indexicality serves as a foundational basis for modeling metapragmatic-centric alignment and provides a dynamic and dialectic complement to metasemantic-centric alignment.

To reiterate, alignment relations represented in Table 3 are not mapped through fixed or demarcated taxonomic differentiations between semantic and pragmatic domains. First, the two alternatively oriented alignments are *communicatively asymmetrically determined* in any specific communicative scenario: the invocation of either metasemantic-centric of metapragmatic alignment is nonetheless governed by contextualization directionality as a metapragmatic criterion (Section 3.2). Within the same or across shared socio-ecological communicative scenes, all rule-like or iconic elements have potential to be contextualized and rendered indexical, but once they are done so, only proper decontextualization and recontextualization together can effectively "switch them back" to a metasemantic-oriented alignment. Second, metasemantic-centric and metapragmatic-centric alignments are *epistemologically equally tenable and pertinent*: metasemantic alignment is not inherently "truer" or more "generalized" than metapragmatic alignment of context and situation, and metapragmatic alignment does not negate the metasemantic values such as in terms of content, structure, logic, or reality. After all, most indexical elements in communication are contextualized from iconic or rule-like elements, and encontextualization typically channels or shifts rather than denies their naturalized or universalized significance, allowing them to function *as if* they were operating in an iconic or rule-like mode. Unlike many theoretical or empirical studies on communicative modeling that relate capacities and functions at the non-meta level, the metasemantic-metapragmatic taxonomic framework promotes a dynamic and dialectic relationship, where metasemantic-centric and metapragmatic centric alignments are alternatively salient yet within both communicative directionality and flexibility embedded. The metasemantic-metapragmatic framework ought to be regarded as a foundational lens for addressing a broad spectrum of alignment challenges in human and machine multimodal communicative modeling[8].

## 5. Conclusion

This paper proposes a metasemantic-metapragmatic framework for grounding and taxonomizing multimodal communicative alignments in human and machine communication. Drawing on pragmatist philosophy and metalinguistic theories, it highlights the dynamic and dialectic interplay between metasemantic and metapragmatic-oriented awareness, with the Principle of Contextualization Directionality underscoring the central role of indexical contextualization in navigating and transitioning between communicative modes and within and across modalities. By grounding communicative alignment in the foundational interplay between metasemantics and metapragmatics, the higher-order, meta-level perspective reframes how other key domains such as intentionality, identity, affect, and ethics, should be conceptualized along with multimodal communication.

## Acknowledgments

Some important ideas for this paper were initially presented at the workshop on Epistemology and AI in the Logic and AI Summit at Zhejiang University. I am particularly grateful to Yang Shen and Xiaoyu Ke for their insightful discussions and thoughtful engagement. I also extend my appreciation to the workshop's participants for their valuable feedback.

---

[8] The four aspects of intentionality, identity, affect, and ethics present an intriguing opportunity to re-examine the dialogue between contemporary multimodal modeling and overarching communicative theories extended from Shannon's information theory. Notably, this includes the six Jakobsonian linguistic functions (Jakobson, 1960) and their contemporary adaptations (Silverstein, 1976; Nakassis, 2023). On the other hand, while Jakobson's framework has significantly influenced the development of contemporary communicative theories in sociolinguistics and linguistic anthropology, certain functions—such as the poetic function in conjunction with the concept of indexicality (Nakassis, 2018)—have been more emphasized over others, leaving room for further contemporary discussions on modeling intentionality, identity, affect, and ethics within multimodal communication.